# A special kind of Casimir plates for making exotic matter

M.Mansooryar
mansooryar@stu.qom.ac.ir

**Abstract**

Herein a process of suitable extraction of energy from vacuum is introduced. By following Ridgely's idea[1], it is proposed some molecular fine-scale plates. Then some properties of them are considered. In fact Casimir effect is viewed in engineering manner. The introduced plates may be applied for support of traversable wormholes (TWs) & warp drives (WDs).

**Introduction**

Herein the basic idea is the introduction of flat Casimir plates in nano scale however changing for engineering reasons, (e.g circle, washer) always remain parallel to themselves. Herein it is followed Ridgely who proposed an approach leading to creating macroscopic amounts of exotic matter (EM) [1]. People who work in field know existence of EM or in other words negative energy (NE) is "almost" necessary for TWs & WDs. The word "almost" is because there is comment about violation of weak energy condition (WEC) with no need to EM such as creation & control of massive scalar or massless scalar [2,3,4,5,6,7,8], nutrino[9,10,11], Brans-Dicke (BD)[12], & so on fields[13,14]. But they don't seem to delete engineering problems which an advanced civilization [15] must do for creation, stabilization & control of TWs or WDs completely. Herein it is shown nano scale plates may be useful. Also some properties are imputed to plates and scene. At last conclusion is if we have enough plates, proper configurations may give us suffician amount of NE.

**Fundamental assumptions:**

It can be briefly mentioned results of [1]:
For two parallel perfect conducting square plates we have:

$$U(r)_{casimir} = \frac{L^2 c \hbar \pi^2}{-720 r^3} \tag{1}$$

$$\underset{r \ll L}{\rightarrow} F(r) = \frac{-\partial U(r)}{\partial r} = \frac{-L^2 c \hbar \pi^2}{240 r^4} \tag{2}$$

$$U_1 = \frac{-L^2 c \hbar \pi^2}{720 r_1^3} \tag{3}$$

$$U_2 = \frac{-L^2 c \hbar \pi^2}{720 r_2^3} \tag{4}$$

$$\Delta U = U_2 - U_1 = \frac{-\pi^2}{720} \hbar c L^2 \left(\frac{1}{r_2^3} - \frac{1}{r_1^3}\right) \tag{5}$$

$$W_{12} = \frac{\pi^2}{720} \hbar c L^2 \left(\frac{1}{r_2^3} - \frac{1}{r_1^3}\right) \tag{6}$$

$$W_\infty = \underset{r_1 \to \infty}{Lim}(W_{12}) \tag{7}$$

$$W_\infty = \frac{L^2 c \hbar \pi^2}{720 r^3} \rightarrow U_\infty = \frac{-L^2 c \hbar \pi^2}{720 r^3} \tag{8}$$

as a result it can be mentioned if zero point field (ZPF) performs positive work in a localized region of spacetime then NE density of the system arises. By the way, there is no difference between inertial properties of EM & common matter.

**Properties of suitable Casimir plates:**

Let's assume a pair of flat structures (plates) which are composed of some conductive quantums. For our purposes a few properties of plates & condition are desired:

A: The shape of plates should be square or rectangular or circular. Square is because easy calculation but circular is for more output of collection (fig 1). As reader knows circle is the extreme area in a flat plate. By the way due to
$U \propto L^2$ increasing the area is very important.

B: According to [16] dependence of Casimir energy to mass is described by:

$$\varepsilon = \frac{1}{a^{d+1} 2^{d+1} \pi^{\frac{(d+1)}{2}} \Gamma(\frac{d+1}{2})} \int_0^\infty dt\, t^d Ln(1 - e^{-2\sqrt{t^2 + \mu^2 a^2}})$$

(9)

$$= -2(\frac{\mu a}{4\pi})^{\frac{d}{2}+1} \frac{1}{a^{d+1}} \sum_{n=1}^\infty \frac{1}{n^{\frac{d}{2}+1}} K_{\frac{d}{2}+1}(2\mu a n)$$

(10)

$$\lambda^2 = \omega^2 - k^2 - \mu^2$$

(11)

here μ denotes mass for the scalar field. So less massive quantums (molecules or atoms), more output of collection.

C: The "plates" should be as much as possible "flat" & a proper topology must be available. So molecules should be as much as possible tiny in particular if there are a few of them.

D: The plates should be light because zero point energy should be extracted more and more & a little of it should move the plates.

E: The plates should be as much as possible thin because more volume of vacuum need to be covered and dug

F: Evidently more conductive plates is better & superconductivity should be considered.

G: All factors leading to friction should be diminished to all forces be conservative. Less wasting is more needed.

H: According to [16] there is a contrary dependence between Casimir force & growing temperature:

$$F^T = \frac{-1}{2\beta} \int \frac{d^d k}{(2\pi)^d} \sum_{n=-\infty}^{n=\infty} \frac{2\kappa_n}{e^{2\kappa_n a} - 1}$$

(12)

$$\kappa_n = [k^2 + (2\pi n/\beta)^2]^{\frac{1}{2}}$$

(13)

$$\beta = 1/kT$$

(14)

So temperature of the surrounding should be low.

**Can we extract NE from vacuum?**

The calculations show with a reachable number of plates it can be produced sufficient NE (dependent to elegance of operation from ~ hundreds to bilions number of plates & distance between them from pico-meters to nano-meters) we can produce needed EM at least for the case reported in[17]. The calculations are under consideration & will be published in my next letter.

---

the below papers are recommended to review:

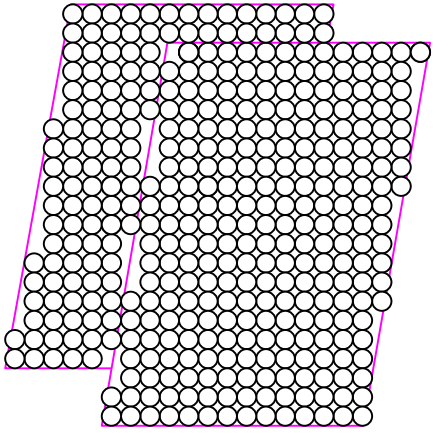 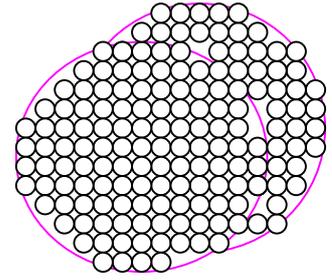

fig1a　　　　　　　　　　　　　　fig1b